\documentclass[conference]{IEEEtran}
\IEEEoverridecommandlockouts
\usepackage{amsmath,graphicx}
\usepackage{amsfonts}
\usepackage{cite}
\usepackage{url}

\title{Aggregation Strategies for Efficient Annotation of Bioacoustic Sound Events Using Active Learning}

\author{\IEEEauthorblockN{Richard Lindholm$^{\S,1}$, Oscar Marklund$^{\S,1}$}
\IEEEauthorblockA{1. Centre for Mathematical Sciences, Faculty of Engineering\\
Lund University, Sweden}
\and
\IEEEauthorblockN{Olof Mogren$^{2,3}$, John Martinsson$^{1,2,3}$}
\IEEEauthorblockA{2. RISE Research Institutes of Sweden}
\IEEEauthorblockA{3. Climate AI Nordics}
}
\begin{document}

\maketitle
\begingroup\renewcommand\thefootnote{\textsection}
\footnotetext{~Equal contribution.}
\endgroup

\begin{abstract}
The vast amounts of audio data collected in Sound Event Detection (SED) applications require efficient annotation strategies to enable supervised learning. Manual labeling is expensive and time-consuming, making Active Learning (AL) a promising approach for reducing annotation effort. We introduce \textit{Top K Entropy}, a novel uncertainty aggregation strategy for AL that prioritizes the most uncertain segments within an audio recording, instead of averaging uncertainty across all segments. This approach enables the selection of entire recordings for annotation, improving efficiency in sparse data scenarios. We compare \textit{Top K Entropy} to random sampling and \textit{Mean Entropy}, and show that fewer labels can lead to the same model performance, particularly in datasets with sparse sound events. Evaluations are conducted on audio mixtures of sound recordings from parks with meerkat, dog, and baby crying sound events, representing real-world bioacoustic monitoring scenarios. Using \textit{Top K Entropy} for active learning, we can achieve comparable performance to training on the fully labeled dataset with only 8\% of the labels. \textit{Top K Entropy} outperforms \textit{Mean Entropy}, suggesting that it is best to let the most uncertain segments represent the uncertainty of an audio file. The findings highlight the potential of AL for scalable annotation in audio and time-series applications, including bioacoustics.
\end{abstract}


\IEEEpeerreviewmaketitle

\section{Introduction}
\label{sec:intro}

The growing biodiversity crisis demands scalable and efficient monitoring techniques to enhance conservation efforts. Bioacoustics, the study of sound in biological contexts, has emerged as a powerful tool for biodiversity monitoring, offering a non-invasive and cost-effective means to collect rich ecological data over large spatial and temporal scales \cite{Stowell2021, browning2017passive}. Analyzing animal vocalizations, environmental sounds, and anthropogenic noise from audio recordings can provide crucial insights into species presence, population dynamics, and ecosystem health. However, the vast amounts of audio data generated by bioacoustic monitoring programs create major challenges in data processing, annotation, and analysis.

Sound Event Detection (SED) plays a vital role in automating the analysis of bioacoustic data, aiming to identify and classify sound events of interest, such as animal vocalizations, within continuous audio streams \cite{Stowell2021}. Supervised deep learning models have achieved remarkable success in SED \cite{mesaros2021sound}, yet their performance hinges on access to large, accurately labeled datasets. The manual annotation of bioacoustic recordings—marking sound events and assigning class labels—is costly, labor-intensive, and constitutes a major bottleneck in developing effective SED models.

\begin{figure}
    \centering
    \includegraphics[width=.4\linewidth]{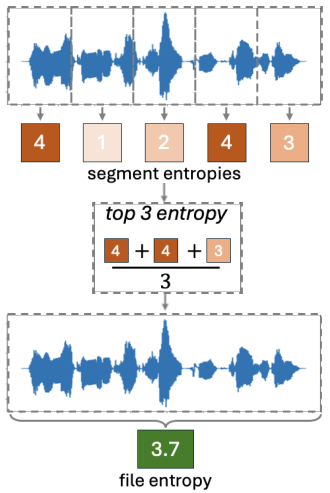}
    \caption{Top $K$ Entropy uncertainty aggregation selects the top $K$ segment entropies (here $K$=3)
obtained from the segments of the file. The resulting uncertainty for the file is the average of the selected segment entropies.}
    \label{fig:enter-label}
\end{figure}
Active Learning (AL) is a promising strategy to mitigate the annotation bottleneck. By selecting the most informative data points for annotation, AL minimizes labeling effort while maintaining high model performance \cite{settles2009active}. While extensively studied in image classification and natural language processing, AL for bioacoustic SED remains underexplored. Existing studies \cite{shuyang2020active, vanosta2023active} have demonstrated its potential but often focus on frame- or segment-based selection rather than the practical scenario where entire audio files must be annotated.


Traditional AL frameworks typically assume independent data points and construct query batches dynamically by combining uncertainty-based selection with diversification strategies~\cite{settles2009active, kirsch2019batchbald, badge2020, ren2021deep}. In contrast, AL for SED operates on segments of audio recordings, which are queried for annotation~\cite{shuyang2020active}. However, in practice, annotation is often performed at the file level, creating a mismatch between segment-based selection and real-world labeling processes. Instead of constructing query batches from independent segments, the challenge shifts to selecting entire recordings by ranking predefined batches of audio segments. This requires new uncertainty aggregation strategies that can effectively translate segment-level uncertainty into file-level scores, making them suitable for batch selection at the recording level.

In this paper, we propose \textit{Top $K$ Entropy}, a novel uncertainty aggregation strategy for active learning in bioacoustic SED. Instead of treating all segments equally, this approach focuses on the most uncertain segments within each file (see figure \ref{fig:enter-label}). Our method is evaluated against other aggregation strategies using a bioacoustic soundscape dataset. The results show that \textit{Top $K$ Entropy} achieves competitive performance with up to 92\% fewer labeled examples. This demonstrates its practical value for large-scale bioacoustic monitoring.

The rest of the paper is organized as follows: Section \ref{sec:al} details our active learning framework, including the SED model and the studied aggregation strategies. Section \ref{sec:evaluation} presents experimental results, highlighting annotation efficiency and generalization. Section \ref{sec:discussion} discusses the broader implications, and Section \ref{sec:conclusions} concludes the paper.

\section{Uncertainty Aggregation for Active Learning in SED}
\label{sec:al}

Our active learning framework is built around a pool-based approach, iteratively refining a segment-based SED model by strategically incorporating new batches of annotated data. 
Our methodology consists of three key components: (1) a segment-based SED model, (2) uncertainty aggregation strategies for selecting entire audio files for annotation, and (3) diversification techniques to improve batch diversity.

\subsection{Segment-based Sound Event Detection Model}
The SED model employed in this study is designed to process short segments in audio files, allowing fine-grained temporal resolution and efficient feature extraction. For feature extraction, we use the YAMNet architecture, a deep convolutional neural network pre-trained on a large-scale audio dataset called AudioSet\cite{gemmeke2017audioset}. YAMNet is known for its ability to capture robust and general-purpose audio representations, making it well suited for transfer learning in bioacoustic domains.

Specifically, each audio file is divided into segments of 0.12 seconds duration, with an overlap of 50\%.  For each segment, a 1024-dimensional feature embedding is extracted.  These embeddings serve as input to a linear classifier that is trained to predict the class label for each segment.  The classifier consists of a single fully connected neural network layer with 4 outputs corresponding to the classes: baby, dog, meerkat and background noise.  A softmax activation function is applied to the output layer, providing probability estimates for each class. The linear classifier is trained using categorical cross-entropy loss and the ADAM optimizer, with hyperparameters tuned for optimal validation performance.

To refine the segment-level predictions and generate coherent sound events, a median filter of kernel size 3 is applied along the temporal dimension of the classifier output. This filtering step smooths the predictions, reduces the effects of transient noise, and enforces temporal consistency. 

After filtering, adjacent segments with the same class prediction are merged into continuous sound events. Consequently, the final output of the SED model consists of a set of detected sound events, each characterized by a class label, start time, and end time.

\subsection{Uncertainty Aggregation Strategies for File Querying}
In our active learning framework, the querying process operates at file level, meaning that entire audio files are selected for annotation in each iteration.  To bridge the gap between segment-level uncertainties and file-level queries, uncertainty aggregation strategies are employed.
These strategies compute a file-level uncertainty score by aggregating segment-level entropies within each unlabeled audio file. The entropy of a segment is quantified by its Shannon entropy, defined as  $E = -\sum_{c} p_c \log_2(p_c)$, where  $p_c$ represents the predicted probability for class $c$. We investigate and compare the following aggregation methods.

\textit{Top $K$ Entropy} is based on the premise that the most uncertain segments provide the most valuable information for improving model performance.
It selects the top $K$ highest segment entropies within a file and calculates their average. By focusing on the $K$ most uncertain segments, this strategy aims to query files that contain a few highly ambiguous segments, potentially indicative of rare or challenging sound events. 
Through empirical evaluation, we found that $K=10$ strikes a balance between selecting sufficiently uncertain segments while avoiding overrepresentation of minor fluctuations in entropy.

Along with this, the following \textit{baseline strategies} were explored in the experiments.
\begin{itemize}
    \item \textit{Mean Entropy} is an aggregation strategy that calculates the average entropy across all segments within a file and uses this as representation of the file-level uncertainty.  It provides a holistic measure of uncertainty, where all segments are considered equal.
    \item \textit{Median Entropy} calculates the median entropy across all segments in a file.
    \item \textit{Mean Event Entropy} uses the mean entropy of all segments which are predicted as events by the model.    
    \item \textit{Random Querying} chooses files at random, without considering entropy. This simulates a scenario where active learning is not applied, serving as a good reference point for all uncertainty aggregation strategies. 
\end{itemize}

In each active learning iteration,  all unlabeled audio files in the training set are ranked based on their computed uncertainty scores from the chosen aggregation strategy. Files with the highest uncertainty scores are considered more informative, and are queried for annotation in a batch of a predefined size.
The annotator is simulated in these experiments by simply revealing the ground truth labels for the queried audio recordings.

\section{Experimental Evaluation and Results}
\label{sec:evaluation}
\subsection{Dataset Generation}
Experiments were conducted using bioacoustic soundscapes to evaluate the performance of different AL strategies.
The datasets provides a controlled and reproducible experimental environment, while closely resembling real-world bioacoustic data. The datasets were generated by mixing foreground sound events with background noise sourced from recordings of park environments \cite{diment2017tut}. The foreground events consisted of vocalizations of three classes: baby cries, \cite{trowitzsch2019nigens}, dog barks \cite{trowitzsch2019nigens}, and meerkat calls \cite{nolasco2022dcase}.  These classes were chosen to represent diverse sound characteristics and event durations. 

datasets were generated with varying characteristics, to assess the robustness of our method under different data conditions. Each dataset was generated using 2 parameters; event ratio $r$, describing the ratio of files containing events, and SNR. Files with events always contain one to three events (chosen uniformly), with equal probability of each class.

The main dataset used was created with an SNR of 0 and $r=0.2$, meaning that 20\% of the audio files contained sound events, while the remaining 80\% consisted solely of background noise. In order to determine if results depend on SNR, we also generated datasets with SNR values of 10dB and -10dB. A dataset with $r=1.0$, where all files contained at least one event, was also created and evaluated. Each of the mentioned datasets comprised of 2500 audio files, with a fixed duration of 10 seconds per file (equivalent of 175 segments).  For each dataset, 80\% of the data was generated for training and 20\% was generated for testing, where the event ratio was kept within each set division.

\subsection{Experimental Setup}
Active learning was simulated using a pool-based setup.
We initialized the AL loop with a randomly selected seed set comprising 0.5\% of the training data and iteratively queried additional unlabeled files for annotation.
The batch sizes were dynamically adjusted throughout the learning process, starting with smaller batches in the early iterations to capture the rapid performance gains of AL, and gradually increasing batch sizes in later iterations.  We compared the performance of \textit{Top $K$ Entropy} with different baseline strategies (\textit{Mean Entropy, Median Entropy, Mean Event Entropy} and \textit{Random Querying}). For each strategy, experiments were repeated for  5  or more random seeds to account for variability in initialization and data sampling. 
Model performance is evaluated using the Intersection over Union (IoU) metric, computed over the entire test set by concatenating data from all testing files.
We refer to this metric as total IoU. 

\subsection{Comparative Analysis of Aggregation Strategies}
\begin{figure}[htb]
    \centering
    \includegraphics[width=\linewidth,trim={0 0 0 2cm},clip]{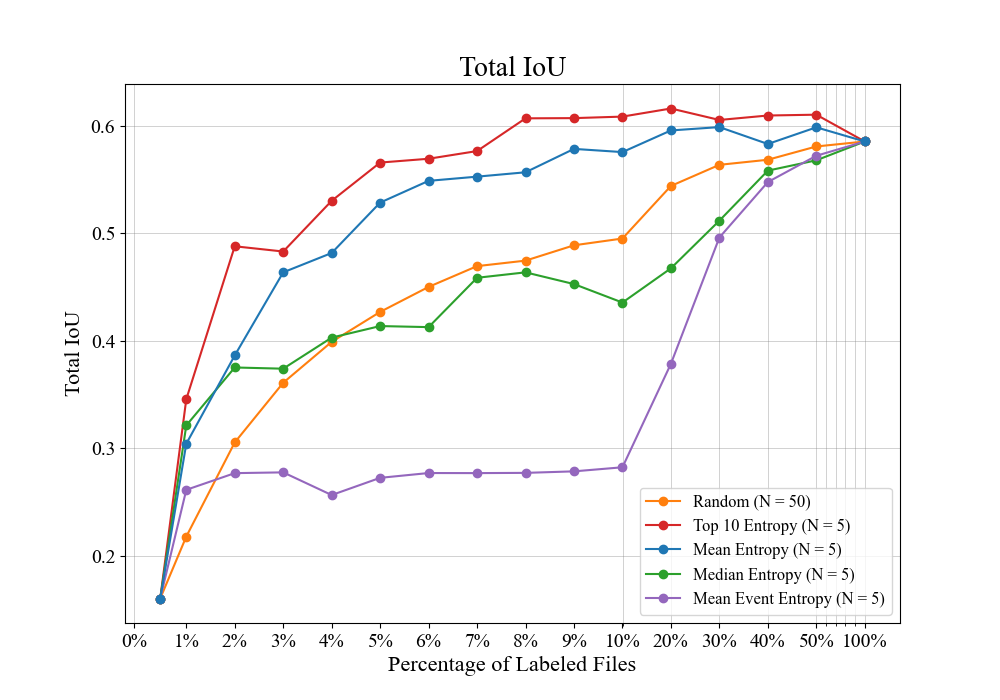}
    \caption{Total IoU performance of different aggregation strategies and the Random Querying baseline, averaged over 5 seeds. \textit{Top K Entropy} achieves comparable IoU to a fully supervised model while reducing annotation effort by 92\%. Results based on data generated with event ratio $r=0.2$ and SNR $=0$.}
    \label{fig:agg_iou}
\end{figure}

Figure \ref{fig:agg_iou} presents a comparison of the total IoU performance for the explored aggregation strategies. Our results indicate that \textit{Top K Entropy} achieves a 92\% reduction in annotation effort, outperforming \textit{Random Querying} across all annotation budgets, thus demonstrating the effectiveness of uncertainty-based selection.
\textit{Top 10 Entropy} outperform the baselines across all annotation budgets, achieving a total IoU comparable to the fully supervised model (trained on 100\% labeled data) with only 8\% of the training data annotated. This translates to a 92\% reduction in annotation effort. \textit{Mean Entropy} also managed to achieve results comparable to a fully supervised model in the early iterations. \\ 
\textit{Median Entropy} does not provide a substantial improvement in annotation efficiency compared to the random baseline. This is not surprising, considering that the majority of segments in each file purely consist of background noise, where the model is both certain and accurate. \textit{Mean Event Entropy} is significantly worse than the random baseline.

The superior performance of \textit{Top 10 Entropy} and \textit{Mean Entropy} suggests that uncertainty-based querying effectively identifies informative audio files for annotation, leading to higher annotation efficiency.

\begin{figure}[htb]
    \centering
    \includegraphics[width=\linewidth,trim={0 0 0 1.7cm},clip]{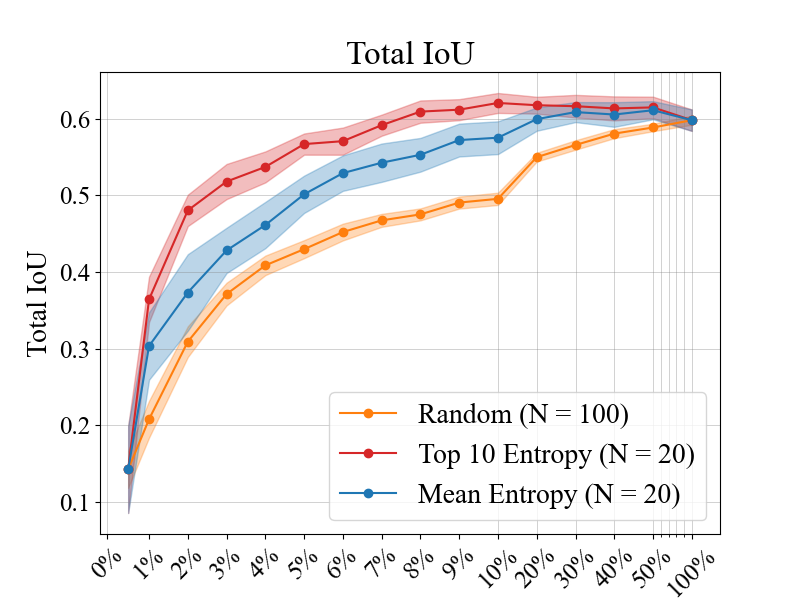} 
    \caption{Total IoU results averaged over 20 seeds (N=20) for \textit{Mean Entropy} and \textit{Top 10 Entropy}. The results for Random Querying baseline are averaged over 20 seeds, with five different initializations for the dataset. Each line is paired with 95\% confidence intervals. Results based on data generated with event ratio $r=0.2$ and SNR $=0$.}
    \label{1.0_merge_confs}
\end{figure}

In figure \ref{1.0_merge_confs}, \textit{Top 10 Entropy} and \textit{Mean Entropy} are further compared to \textit{Random Querying}, with 95\% confidence intervals. This experiment was run for 20 different seeds to get better approximations of average performance and variance. These results clearly show the superiority of \textit{Top 10 Entropy}. The effectiveness of \textit{Top $K$ Entropy} highlights the importance of focusing on the most ambiguous segments within audio files, as these segments likely contain novel or challenging sound events that contribute most significantly to model learning. Given that \textit{Top $K$ Entropy} outperforms \textit{Mean Entropy}, it suggests that querying files with the highest peaks in uncertainty is more effective than querying those with the highest average uncertainty.

Diversification strategies were evaluated (Farthest Traversal \cite{shuyang2020active}, and Random Selection), and they improved the results for the baseline strategies slightly. The \textit{Top $K$ Entropy} strategy, on the other hand, appears to inherently select diverse batches, as its performance remained unaffected by additional diversification techniques. The \textit{Top $K$ Entropy} strategy consistently had the best scores across all experiments.

\subsection{Exploring Top $K$ Entropy}
\begin{figure}[htb]
    \centering
    \includegraphics[width=\linewidth,trim={0 0 0 1.6cm},clip]{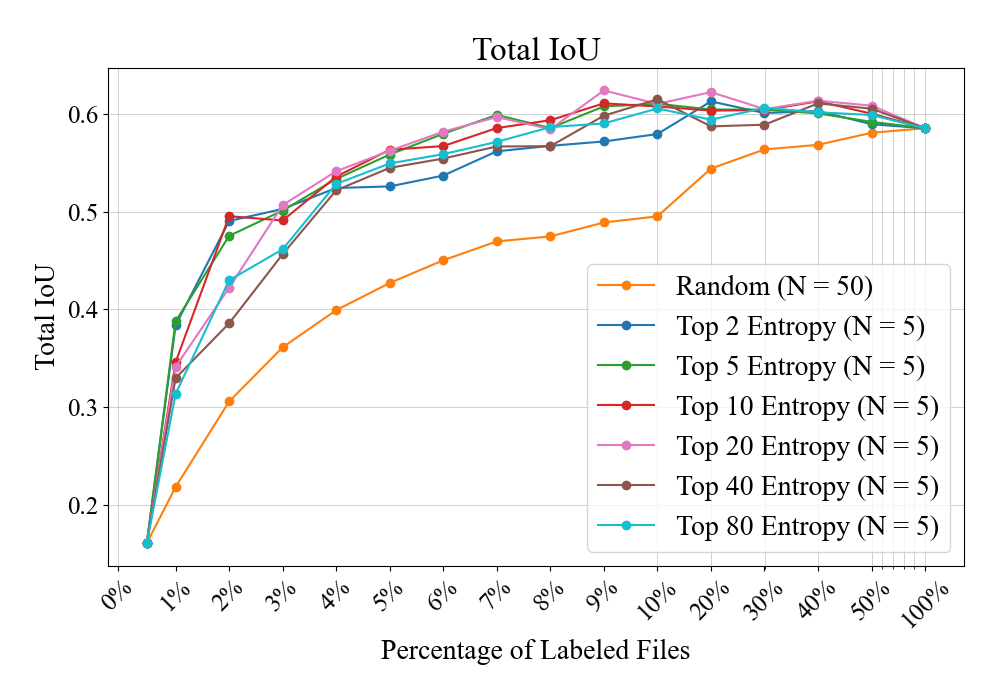}
    \caption{Total IoU performance averaged over 5 seeds (N=5), for 6 different fixed values of $K$ for the querying strategy \textit{Top $K$ Entropy}) along with the baseline strategy. Results based on data generated with event ratio $r=0.2$ and SNR $=0$.}
    \label{fig:x_top_x}
\end{figure}

$K$ was set to $10$ in the initial testing of the \textit{Top $K$ Entropy} strategy. In figure \ref{fig:x_top_x} a comparison between different values of $K$ is presented. \\
All tested values of \textit{$K$} outperform \textit{Random Querying}, but differences between them are hard to distinguish. Smaller values of \textit{$K$} seem to perform slightly better for low annotation budgets. These results show that the \textit{Top $K$ Entropy} is robust with respect to the selection of $K$. Previously, in figure \ref{1.0_merge_confs}, we observed that the maximum value of $K=175$ (which corresponds to \textit{Mean Entropy}) performs worse compared to $K=10$. This suggests that there is an upper limit for a suitable value of $K$.

\subsection{Generalization - SNR}
The sensitivity of models to varying SNR was evaluated using SNR $=10$ and SNR $=-10$, showing an improvement and a degradation in performance, respectively.
%
%
%
In both cases, the active learning framework is more annotation efficient compared to the baseline. For SNR $= -10$ the relative improvement from the baseline is much higher, compared to when the data is mixed with a higher SNR. This suggests that the relative gain of our approach is bigger given a more challenging SED task. These results also showcase robustness to noise power variability, beneficial for real-world applications where the SNR will fluctuate. 

\subsection{Generalization - Event Ratio}
To evaluate the impact that event ratio has on active learning performance, a dataset consisting solely of files containing events was used.
%
%
 The \textit{Top 10 Entropy} strategy consistently outperformed the baseline, indicating robustness to changes in event frequency. However, in this domain the \textit{Mean Entropy} strategy did not outperform the random baseline, suggesting it may be better suited for datasets with fewer events. Overall, the active learning strategies demonstrated greater effectiveness when fewer events were present in the data, suggesting that active learning is most beneficial when events are scarce.

\section{Discussion}
\label{sec:discussion}

Our results demonstrate the effectiveness of the \textit{Top $K$ Entropy} aggregation strategy for active learning for bioacoustic SED.  The strategy consistently demonstrates superior performance, suggesting that focusing on the most uncertain segments within audio files is crucial for efficient learning.  The significant reduction in labeled data required to achieve high performance underscores the practical value of the proposed strategy in this domain.

The analysis of queried files reveals that successful AL strategies tend to address class imbalance and prioritize files with sound events. While diversification techniques showed some potential for further improvement, particularly Farthest Traversal with AudioMAE \cite{audiomaehuang2023masked} embeddings, the gains were modest in our experiments. This could be due to the effectiveness of \textit{Top $K$ Entropy} in already selecting a relatively diverse set of informative files.  Further investigation into more sophisticated diversification strategies and their interaction with different uncertainty aggregation methods is warranted.

The generalization results across SNR levels further underscore the robustness and practical applicability of the proposed approach. The baseline approaches suffer more from a lower SNR than \textit{Top $K$ Entropy}, suggesting that the approach is robust and stable in different environments. This is important for the deployment of bioacoustic SED systems in diverse ecological settings and under varying environmental conditions.

The generalization results for the dataset containing only files with events further highlights the potential of \textit{Top K Entropy}. As the event ratio increases, the reward for identifying files containing events diminishes, while prioritizing files with a higher number of events or greater event variability becomes more important. The results indicate that \textit{Top K Entropy} adapted better to this change in data distribution than the other baselines.

The success of \textit{Top K Entropy} suggests that prioritizing the most uncertain segments within a file provides a better representation of file-level uncertainty than traditional averaging methods. This aligns with the broader active learning literature, which highlights that focusing on high-entropy regions can accelerate model convergence. Future studies should explore whether this principle extends to other time-series domains, such as speech recognition and medical diagnostics.

Limitations of this study include the use of soundscapes mixed with known sound event classes, which, while designed to mimic real-world data, may not fully capture the complexities of natural bioacoustic recordings. 
Future research should validate our findings using real-world bioacoustic datasets and explore how self-supervised learning methods, such as AudioMAE or wav2vec, can enhance uncertainty estimation in active learning frameworks.
Furthermore, our study does not fully examine the impact of class imbalance on \textit{Top K Entropy's} effectiveness. If rare events consistently exhibit high entropy, they may be overrepresented in queries, potentially biasing the training process. Future research should investigate adaptive sampling strategies that balance uncertainty with class distribution awareness.

\section{Conclusions}
\label{sec:conclusions}

This research provides an evaluation of uncertainty aggregation strategies in active learning for efficient annotation in bioacoustic sound event detection.  Our findings demonstrate that uncertainty-based AL, and particularly the \textit{Top $K$ Entropy} aggregation strategy, offers a powerful approach to drastically reduce annotation effort while maintaining high SED performance.
By prioritizing the most uncertain segments, the \textit{Top $K$ Entropy} strategy reduces annotation requirements by up to 92\% while maintaining model performance, demonstrating its potential for scalable bioacoustic monitoring.
The robustness of the proposed approach across SNR variations and varying event ratios further underscores its practical applicability for real-world bioacoustic monitoring.

Future research directions include exploring dynamic adaptation of the $K$ parameter in \textit{Top $K$ Entropy} to optimize performance across different data distributions and annotation budgets.  Investigating more advanced diversification techniques and their interplay with uncertainty aggregation methods could also lead to further improvements in annotation efficiency.  Furthermore, evaluating AL strategies on diverse real-world bioacoustic datasets and exploring the integration of AL with citizen science initiatives could pave the way for scalable and cost-effective biodiversity monitoring solutions.  By significantly reducing the annotation bottleneck, active learning has the potential to democratize bioacoustic data analysis, empowering researchers and conservation practitioners to leverage the vast amounts of acoustic information for effective biodiversity monitoring and conservation action.

\section{Acknowledgements}
This work was supported by Swedish Foundation for Strategic Research, FID20-0028.

\vfill

\bibliographystyle{IEEEtran}
\bibliography{refs}

\end{document}